\documentclass[a4paper,fleqn]{cas-dc}
\usepackage{dblfnote}
\usepackage[authoryear,longnamesfirst]{natbib}
\usepackage{xcolor}
\usepackage{algorithm}
\usepackage{algorithmic}

\usepackage{graphicx} 
\usepackage{lipsum} 
\usepackage{float}
\usepackage{placeins}

\hypersetup{
	colorlinks=true,
	linkcolor=blue,
	urlcolor=blue,
	linktoc=all
}
\def\tsc#1{\csdef{#1}{\textsc{\lowercase{#1}}\xspace}}
\tsc{WGM}
\tsc{QE}
\tsc{EP}
\tsc{PMS}
\tsc{BEC}
\tsc{DE}
\begin{document}
	\let\WriteBookmarks\relax
	\def\floatpagepagefraction{1}
	\def\textpagefraction{.001}
	
	% Short title
	\shorttitle{ STL-FFT-STFT-TCN-LSTM }
	
	% Short author这个位置是作者名字的缩写：
	\shortauthors{ }
	
	% Main title of the paper
	\title[mode = title]{STL-FFT-STFT-TCN-LSTM: An Effective Wave Height High Accuracy Prediction Model Fusing Time-Frequency Domain Features}  
	
	% Title footnote mark
	% eg: \tnotemark[1]
	% \tnotemark[<tnote number>] 
	% \tnotemark[1,2]
	
	% \tnotetext[1]{This document is the results of the research project funded by the National Science Foundation.}
	% \tnotetext[2]{The second title footnote which is a longer text matter to fill through the whole text width and overflow into another line in the footnotes area of the first page.}

        \author[1,3]{ Huipeng Liu }[orcid= U2006211] 
        % \ead{ changlu@tju.edu.cn } 
        
        % \credit{Conceptualization of this study, Methodology, Software}
        
        \author[2,3]{ Zhichao Zhu }
        
        \author[2,3]{ Yuan Zhou }
        % \ead{rishi@sayahna.org}
        % \ead[URL]{www.sayahna.org}
        \credit{Data curation, Writing - Original draft preparation}
        
        \author[1,3]{ Changlu Zhou }
        \cormark[1]
        % \ead{karl@freefriends.org} 
        % \ead[URL]{www.tug.org}
        
        \address[1]{ Tianjin University, Tianjin 30072, China }
        \address[2]{ }
        \address[3]{ }
        
        \cortext[1]{  zhouyuan@tju.edu.cn } 
        % \cortext[2]{Principal corresponding author}
    
	% \author[1,3]{ Huipeng Liu }[type=editor,
	% auid=000,bioid=2,
	% % prefix=Sir, 
	% role= , 
	% orcid= ]
	% \cormark[1] 
	% \fnmark[1] 
	% \ead{ } 
	% % \ead[url]{www.cvr.cc,www.tug.org.in}
	% \credit{Conceptualization of this study, Methodology, Software}
	
	% \author[2,3]{ Zhichao Zhu }[style=chinese]
	
	% \author[2,3]{  }
	% \fnmark[2] 
	% % \ead{rishi@sayahna.org}
	% % \ead[URL]{www.sayahna.org}
	% \credit{Data curation, Writing - Original draft preparation}
	
	% \author[1,3]{ }
	% \fnmark[1,3]
	% % \ead{karl@freefriends.org} 
	% % \ead[URL]{www.tug.org}
	
	% \address[1]{ }
	% \address[2]{ }
	% \address[3]{ }
	
	% \cortext[1]{Corresponding author} 
	% \cortext[2]{Principal corresponding author} 
	
	\begin{abstract}
		As the consumption of traditional energy sources intensifies and their adverse environmental impacts become more pronounced, wave energy stands out as a highly promising member of the renewable energy family due to its high energy density, stability, widespread distribution, and environmental friendliness. The key to its development lies in the precise prediction of Significant Wave Height (WVHT). However, wave energy signals exhibit strong nonlinearity, abrupt changes, multi-scale periodicity, data sparsity, and high-frequency noise interference; additionally, physical models for wave energy prediction incur extremely high computational costs. To address these challenges, this study proposes a hybrid model combining STL-FFT-STFT-TCN-LSTM. This model exploits the Seasonal-Trend Decomposition Procedure based on Loess (STL), Fast Fourier Transform (FFT), Short-Time Fourier Transform (STFT), Temporal Convolutional Network (TCN), and Long Short-Term Memory (LSTM) technologies. The model aims to optimize multi-scale feature fusion, capture extreme wave heights, and address issues related to high-frequency noise and periodic signals, thereby achieving efficient and accurate prediction of significant wave height. Experiments were conducted using hourly data from NOAA Station 41008 and 41047 spanning 2019 to 2022. The results showed that compared with other single models and hybrid models, the STL-FFT-STFT-TCN-LSTM model achieved significantly higher prediction accuracy in capturing extreme wave heights and suppressing high-frequency noise, with MAE reduced by 15.8\%-40.5\%, SMAPE reduced by 8.3\%-20.3\%, and R² increased by 1.31\%-2.9\%; in ablation experiments, the model also demonstrated the indispensability of each component step, validating its superiority in multi-scale feature fusion.
	\end{abstract}
	
	\maketitle
	
	\section{Introduction}
	In recent years, global temperatures have continued to rise, glaciers have accelerated their melting, sea levels have steadily increased, the ozone layer has experienced severe depletion, and atmospheric pollution has intensified (\cite{IEA2023CO2Emissions};\cite{IEA2025GlobalEnergyReview}). In response, reducing reliance on highly polluting energy sources has become an urgent global priority, attracting significant attention from the international community and academic researchers. Among clean renewable energy sources, wave energy is notable for its higher energy density and greater predictability compared with solar and wind power (\cite{Guo2021Review}), making it a prominent focus of current research. In this context, significant wave height (WVHT) is a critical parameter for wave energy forecasting, playing an essential role in marine engineering design, maritime safety, and early warning systems for marine disasters. WVHT, which is defined as the average height of the highest one-third of waves within a given period, serves as a key indicator for characterizing wave conditions. However, the accurate prediction of WVHT is challenged by the complex and dynamic nature of marine environments and the chaotic characteristics of wave data. \cite{Sithara2025MachineLearning} noted that the main forecasting difficulties arise from nonlinear interactions among wind speed, ocean currents, and topography, as well as the high volatility and non-stationarity of wave measurements. These factors complicate model development and often result in forecasting errors. Over recent decades, researchers have developed four mainstream approaches to address these challenges: numerical wave models, classical time series models, advanced nonlinear models based on intelligent technologies, and hybrid models.
	
	Numerical wave models simulate wave propagation based on energy balance equations (EBEs) that adhere to physical principles. According to the various components of the source function within EBEs, numerical wave models are categorised into three generations (\cite{Jansen2000}). First-generation wave models considered simple wind fields without dominant nonlinear interactions or energy dissipation. Second-generation models incorporated variable wind fields featuring nonlinear interactions. Third-generation wave models were further developed by employing source functions with no a priori constraints on spectral shape (\cite{Tolman1996}). The Wave Model (WAM) was the first operational third-generation model, suitable for global-scale forecasting. Subsequently developed Simulating Waves Nearshore (SWAN) and WAVEWATCH III (WW3) optimised simulation capabilities for nearshore and complex coastal environments. These models rely on high-resolution wind field data to achieve long-term WVHT forecasting, yet their substantial memory requirements, computational time demands, and sensitivity to wind field errors constrain their real-time application (\cite{Booij1999ThirdGenerationWaveModel}; \cite{Tolman1991ThirdGenerationModel}). Machine learning approaches utilising observed wave data offer a complementary pathway for real-time wave forecasting. These methods feature relatively straightforward modelling processes that bypass wind-to-wave conversion, thereby avoiding substantial computational costs.
	
	Early wave forecasting research employing machine learning utilised classical time series models such as autoregressive (AR), autoregressive moving average (ARMA), and autoregressive integrated moving average (ARIMA) models. These approaches predict wave heights using historical data with high computational efficiency. \cite{Spanos1983ARMA} employed ARMA models to simulate the time series of a single wave over a short, stationary period. \cite{SOARES1996277} utilised AR models to describe effective wave height time series at various locations along the Portuguese coast. Subsequently, \cite{GUEDESSOARES2000297} extended AR models to bivariate sequences comprising effective wave height and mean period.
	
	However, the linearity and stationarity assumptions of classical time series models necessitate complex preprocessing for non-stationary, non-linear WVHT sequences (\cite{Ban2023ShortTermPrediction}). In contrast, intelligent technology-based approaches enable nonlinear modelling without requiring prior knowledge of the relationship between input and output variables. Consequently, they represent more generalised and flexible modelling tools for nonlinear waves. In recent years, machine learning algorithms have demonstrated outstanding performance in WVHT forecasting due to their nonlinear modelling capabilities. While artificial neural networks (ANN) (\cite{Deo1998RealTimeWaveForecasting}), extreme learning machines (ELM) (\cite{Zhu2018SpatiotemporalVisualSaliency}), and support vector machines (SVM) (\cite{Mahjoobi2009Prediction}) have been widely applied, convolutional neural networks (CNN) and long short-term memory networks (LSTM) (\cite{Feng2020PreliminaryStudy}; \cite{Fan2020NovelModel}) have further enhanced accuracy. CNN effectively extract spatially correlated features within wave signals through local receptive fields and weight sharing mechanisms, demonstrating particular efficacy when integrating multi-source ocean environment data such as wind speed and atmospheric pressure with wave height coupling relationships. \cite{Yang2022TemporalConvolutionalNetwork} employed CNN to extract features from imaged wave echoes and estimate wave height based on X-band radar data, validating their capability to capture local wave signal characteristics; LSTM networks resolve the vanishing gradient issue inherent in traditional recurrent neural networks through gating mechanisms. They excel at capturing long-term dependencies in time series, offering superior modelling capabilities for periodic fluctuations and abrupt changes within wave data. \cite{FAN2020107298} proposed an LSTM-based effective wave height prediction model, which achieved lower prediction errors than traditional machine learning models for short-term forecasts by learning the dynamic patterns of historical wave height sequences; \cite{Klemm2023Predicting} further applied LSTM to future wave height forecasting, confirming their efficacy in capturing long-term trends in wind-wave coupling. Although CNN and LSTM each possess advantages in handling non-linear and temporal features, the pervasive non-stationarity inherent in ocean waves (\cite{Huang1999}) remains a challenge for both. While CNN excel at extracting local features, they struggle to directly model long-term temporal dependencies; conversely, LSTM, though adept at temporal modelling, exhibit insufficient sensitivity to high-frequency noise and transient fluctuations (\cite{DEKA201232}). Particularly in long-range forecasting, a single model remains inadequate for integrating multi-scale features. Indeed, the nonlinearity and non-stationarity of wave signals often involve multi-mode superposition. The feature extraction capabilities of a single deep learning model are inherently limited, making it difficult to comprehensively capture all key features within complex marine environments. To overcome these limitations, hybrid models have gradually emerged as a research focus.

	Wavelet-based models have been proposed for wave forecasting, as wavelet analysis is effective for non-stationary time series. \cite{ozger2010} introduced the Wavelet Fuzzy Logic (WFL) model by incorporating the wavelet transform into fuzzy logic (FL) methodology. \cite{DEKA201232} developed the wavelet neural network (WLNN) model by combining ANN with wavelet transformation. In their studies (\cite{DEKA201232}; \cite{ozger2010}), wavelet techniques were employed to decompose broadband effective wave height time series into several narrowband components, which were then used as inputs for ANN and FL models, respectively. Prediction results indicated that hybrid models outperformed single models such as ARMA, FL, and ANN. However, wavelet-based hybrid models also exhibit limitations. In essence, the wavelet transform is a linear and non-stationary method that represents signals through linear combinations of wavelet basis functions. Consequently, it may prove unsuitable for non-linear data (\cite{Wu2009EnsembleEM}). Hybrid forecasting models require more effective decomposition techniques to address non-linearity and non-stationarity in real time. An adaptive, data-driven technique known as Empirical Mode Decomposition (EMD) proves highly effective in analysing non-linear and non-stationary datasets (\cite{Huang1998TheEM}). It essentially functions as a dual-stage filter (\cite{Flandrin2004}), decomposing broadband complex signals into relatively simple components with distinct temporal scales. Combining wavelet decomposition or EMD with neural networks can significantly enhance accuracy, although EMD exhibits lower efficiency when processing long sequences (\cite{Seemanth2016SensitivityAnalysis}). \cite{Duan2016HybridEMDSVR} integrated EMD with Support Vector Regression (SVR), substantially reducing prediction errors but maintaining low efficiency for lengthy sequences. \cite{Yang2021NovelHybridModel} proposed the STL-CNN-PE model by combining Loess-based Seasonal-Trend decomposition (STL), CNN, and Position Encoding (PE). In both nearshore and offshore environments, this model significantly outperformed single models and the EMD-SVR model in terms of performance, prediction error, efficiency, and correlation; however, it did not substantially address high-frequency noise and extreme wave height issues.
	
	To address the shortcomings of existing methods, this study proposes a novel hybrid model STL-FFT-STFT-TCN-LSTM rather than other algorithms. Its core lies in precisely matching the key challenges of WVHT forecasting with the characteristics of each algorithm, employing a comprehensive 'decomposition-feature extraction-modelling' workflow to overcome the limitations of traditional approaches. WVHT signals encompass deterministic components such as trends and seasonality, alongside dynamic elements like extreme wave heights and high-frequency noise. Failure to separate these components leads to model learning confusion. STL, based on LOESS local weighted regression, exhibits strong robustness to outliers, enabling precise extraction of trend and seasonal components. This allows residuals to focus solely on extreme wave heights and high-frequency noise. In contrast, EMD exhibits low efficiency when processing long sequences and is sensitive to extreme values. Wavelet decomposition, being a linear method, struggles to accommodate the strong non-linearity inherent in wave signals. Consequently, STL is selected over other decomposition methods. The frequency characteristics of WVHT signals exhibit both periodicity and time-varying transient properties, which cannot be comprehensively captured by single-frequency domain methods. FFT performs a global frequency domain transformation on residual components, extracting stable global spectral features. This provides a foundation for capturing anomalous spectral patterns corresponding to high-frequency noise and extreme wave heights. STFT generates time-frequency matrices via sliding windows to capture time-varying signal characteristics, tracking temporal nodes and trend shifts arising from data discontinuities during non-linear dynamics. This compensates for FFT limitations, as standalone FFT cannot process time-varying frequencies in non-stationary signals. Thus, the two methods complementarily cover frequency domain features. Compared to other frequency domain approaches, traditional wavelet transforms struggle to accommodate strong non-linearity, while Hilbert - Huang transform (HHT) is prone to modal aliasing and exhibits low efficiency (\cite{Cao2016Modal}), while a standalone STFT lacks a global spectral reference. Consequently, the combination of FFT and STFT better accommodates the characteristics of wave signals. Multi-source features exhibit complex interdependencies, and WVHT possesses long-term dependencies that a single network struggles to accommodate. TCN fuses multi-scale features through dilated causal convolutions, uncovering inter-feature correlations to generate high-dimensional abstract representations, thereby addressing the shortcomings of multi-source feature fusion. LSTM captures long-term dependencies through gating mechanisms, circumventing the vanishing gradient problem in traditional RNNs and resolving long-period modelling failures. Single CNNs cannot model long-term dependencies, while Transformers exhibit high computational complexity, low efficiency with massive hourly-scale data, and sensitivity to high-frequency noise. Hence, TCN and LSTM are chosen for collaborative processing. The ablation experiments in Section 4.4 further validate the necessity of this hybrid strategy. Removing any module from STL, FFT, or STFT—or employing only STL—results in a significant increase in MAE, confirming the irreplaceable nature of each algorithm.

	The remainder of this paper is organised as follows: Section 2 introduces the theoretical formulas for STL, FFT, STFT, TCN, and LSTM; Section 3 discusses data sources, distribution, preprocessing, and model architecture; Section 4 introduces model performance evaluation criteria, analyzing the performance of the STL-FFT-STFT-TCN-LSTM model, single machine learning models, and hybrid models across the same time scale and sea area from multiple perspectives, and conducting ablation experiments on the STL-FFT-STFT-TCN-LSTM model; Section 5 summarizes the content and work of this paper and outlines future directions.

	\section{Methodology}
	\subsection{Seasonal-trend decomposition procedure based on loess (STL)}
	The STL algorithm was formally proposed by \cite{Cleveland1990STL} and is a time series decomposition method based on local weighted regression (LOESS). Compared with traditional decomposition methods, a significant advantage of STL is its stronger robustness to time series containing outliers—this enables it to effectively handle transient abnormal behaviour in the data. The algorithm decomposes the time series into three components using an additive model: \( Y_t = T_t + S_t + R_t \), where \( T_t \) denotes the trend term, reflecting the long-term evolutionary trend of the time series; \( S_t \) represents the seasonal term, capturing periodic fluctuations;  \( R_t \) is the residual term, containing sudden change signals of extreme wave heights and periodic signals of high-frequency noise.
	
	The implementation of STL relies on nested iterations. The inner loop iterates to update the trend and seasonal components, extracting the long-term trend and periodic characteristics of the signal; the outer loop calculates robust weights after each inner loop iteration, which are used in the next round of inner loops to reduce the impact of outliers on the trend and seasonal components. The pseudocode for the STL algorithm is shown in Algorithm 1.
	
	\begin{algorithm}
		\caption{STL Algorithm}
		\begin{algorithmic}[1]
			\REQUIRE A time series $X = \{X_t\}$
			\ENSURE Trend component $T_t$, seasonal component $S_t$, and residual component $R_t$
			\STATE Initialize $T_t^0 = 0$
			\FOR{Each outer-loop iteration $k$ until convergence is achieved}
			\FOR{Each inner-loop iteration until convergence is achieved}
			\STATE Step 1: Detrending: $X_t' = X_t - T_t^{k}$
			\STATE Step 2: Periodic subsequence smoothing: Apply a Loess smoother to each periodic subsequence to obtain the temporary seasonal sequence $\tilde{S}_t^{k+1}$
			\STATE Step 3: Low-pass filtering: Implement a low-pass filter with smoothing on $\tilde{S}_t^{k+1}$ to identify the residual trend $\tilde{T}_t^{k+1}$
			\STATE Step 4: Detrending the smoothed periodic subsequence: $S_t^{k+1} = \tilde{S}_t^{k+1} - \tilde{T}_t^{k+1}$
			\STATE Step 5: Removal of seasonal effects: $X_t'' = X_t - S_t^{k+1}$
			\STATE Step 6: Trend smoothing: Apply a Loess smoother to the deseasonalized sequence to obtain $T_t^{k+1}$
			\ENDFOR
			\STATE Compute residuals: $R_t^{k+1} = X_t - S_t^{k+1} - T_t^{k+1}$
			\STATE Update the trend and seasonal components: $T_t^{k+1} = T_t^{k+1}$, $S_t^{k+1} = S_t^{k+1}$
			\STATE Check for convergence:
			\IF{Converged}
			\STATE Output the trend $T_t$, seasonality $S_t$, and residual $R_t$
			\ELSE
			\STATE Return to the outer loop to continue iterations
			\ENDIF
			\ENDFOR
		\end{algorithmic}
	\end{algorithm}
	
	\subsection{Fast Fourier Transform (FFT)}
	
	FFT is used in modern signal processing. It is an algorithm designed to efficiently calculate the discrete Fourier transform (DFT) and its inverse transform. The direct time complexity of DFT is is \(O(N^2)\), while FFT can reduce it to \(O(N \log N)\), significantly improving computational efficiency(\cite{Yuan2016AreaEfficient}). The core idea of FFT is the divide-and-conquer strategy. Its most commonly used implementation is the Cooley-Tukey algorithm, which splits the DFT of length \(N\) into multiple smaller-scale DFTs through recursive decomposition, ultimately simplifying complex computations into a series of simple subproblems. Let the signal length be \(N = 2^m\) (where \(m\) is an integer); the input signal X[n] can be divided into odd-indexed and even-indexed sequences, the DFT can be decomposed into two smaller DFTs of length \(N/2\), which are then combined for computation, as in formula (1) and (2).  
	\begin{equation}
		X[k] = X_{\text{even}}[k] + e^{-\frac{2\pi i k}{N}} X_{\text{odd}}[k]
	\end{equation}
	\begin{equation}
		X\left[k + \frac{N}{2}\right] = X_{\text{even}}[k] - e^{-\frac{2\pi i k}{N}} X_{\text{odd}}[k]
	\end{equation}
	Here, \( e^{-\frac{2\pi i k}{N}} \) is referred to as the twiddle factor. Through recursive decomposition, the DFT calculation can ultimately be simplified to the solution of multiple smaller subproblems. The structure of the FFT module is shown in Algorithm 2.
	
	% % FFT
	\begin{algorithm}[!h]
		\caption{FFT Algorithm}
		\label{alg:FFT}
		\renewcommand{\algorithmicrequire}{\textbf{Input:}}
		\renewcommand{\algorithmicensure}{\textbf{Output:}}
		
		\begin{algorithmic}[1]
			\REQUIRE $x$ (ARRAY OF COMPLEX) \hfill (Input signal) %% input
			\ENSURE $X$ (ARRAY OF COMPLEX) \hfill (FFT output) %% output
			
			\STATE $N \gets \text{LENGTH}(x)$
			\IF{$N = 1$}
			\RETURN $x$
			\ENDIF
			
			\STATE let even $\gets \text{FFT}(x[0], x[2], \ldots, x[N-2])$  \hfill (Take the even indices)
			\STATE let odd $\gets \text{FFT}(x[1], x[3], \ldots, x[N-1])$   \hfill (Take the odd indices)
			
			\FOR{$k \gets 0 \text{ to } N/2 - 1$}
			\STATE $t \gets \text{EXP}(-2 \cdot \pi \cdot i \cdot k / N) \cdot \text{odd}[k]$
			\STATE $X[k] \gets \text{even}[k] + t$
			\STATE $X[k + N/2] \gets \text{even}[k] - t$
			\ENDFOR
			
			\RETURN $X$
		\end{algorithmic}
	\end{algorithm}
	
	\subsection{Short-Time Fourier Transform (STFT)}
	
	STFT is a time-frequency analysis method used to process non-stationary signals. It applies a sliding window function on the time axis and performs a Fourier transform on the signal within the window to generate a two-dimensional time-frequency representation, thereby capturing the time-varying frequency characteristics of the signal (\cite{Yuegang2007NonStationarySignals}).
	
	For discrete-time STFT, the mathematical definition is formula (3).
	\begin{equation}
		X(m, k) = \sum_{n=0}^{N-1} x[n + mR] \omega[n] e^{-j\frac{2\pi kn}{N}}
	\end{equation}
	where \( x[n] \) is the original signal, \( \omega[n] \) is the window function, \( N \) is the length of Fourier transform per frame, \( m \) is the frame index, and \( R \) is the frame shift.

	\subsection{Temporal Convolutional Network (TCN)}
	TCN is a neural network architecture specifically designed for sequence modelling, with its core lying in the combination of causal convolution, dilated convolution, and residual modules to achieve the integration of multi-scale features and long-range dependency modelling (\cite{Yang2022TemporalConvolutionalNetwork}). Among these, causal convolution is the foundation of TCN, ensuring that the output at each time step depends solely on historical information and avoids leakage of future information; dilated convolution expands the receptive field by inserting gaps (controlled by the dilation rate d) between elements of the convolution kernel; residual blocks mitigate the vanishing gradient problem in deep networks through information shortcutting. Each residual block contains two layers of dilated convolutions. The formulas of three modules are shown in (4), (5) and
	(6).
	\begin{equation}
		y(t) = \sum_{i=0}^{k-1} f(i) \cdot x(t - i)
	\end{equation}
	
	\begin{equation}
		y(t) = \sum_{i=0}^{k-1} f(i) \cdot x(t - d \cdot i)
	\end{equation}
	
	\begin{equation}
		Y = \text{ReLU}\left( \text{Dropout}\left( \text{Conv}_2 \left( \text{ReLU}\left( \text{Dropout}\left( \text{Conv}_1 (X) \right) \right) \right) \right) \right) + X
	\end{equation}
	
	In this context, \( y(t) \) denotes the convolution output at time step \( t \); \( k \) represents the size of the convolution kernel; \( f(i) \) indicates the \( i \)-th weight parameter of the convolution kernel; \( x(t-i) \) stands for the value of the input sequence at time \( t-i \); \( d \) is the dilation factor; \( X \) denotes the input of the residual block; and \( Y \) represents the output of the residual block.  
	
	\subsection{Long Short-Term Memory (LSTM)}
	LSTM addresses the vanishing gradient problem in traditional recurrent neural networks (RNNs) by introducing a gating mechanism, enabling the modeling of long-term dependencies in sequences (\cite{FAN2020107298}). Its core innovation lies in the design of the cell state and three gates that control information flow: the forget gate, input gate, and output gate. The forget gate determines which historical information in the cell state should be discarded; the input gate controls whether new information from the current input is stored in the cell state; the cell state is updated to integrate historical memory with candidate information derived from the current input, forming a new cell state; and the output gate determines which information from the cell state should be output. The formulas of three modules are shown in (7), (8), (9), (10), (11) and (12).
	
	LSTM updates the cell state via a gating mechanism, taking the input \( x_t \) at time step \( t \), the hidden state \( h_{t-1} \), and the cell state \( c_{t-1} \) from the previous time step as inputs. Herein, \( f_t \), \( i_t \), and \( o_t \) denote the forget gate, input gate, and output gate, respectively;  \( \tilde{C}_t\) represents the candidate cell state. \( h_t \) and \( h_{t-1} \) denote the hidden states of the forward and backward LSTMs, respectively. \( C_t \) and \( C_{t-1} \) denote the forward and backward cell states at time \( t \). \( \boldsymbol{x}_t \) is the input vector at the current time step; \( \boldsymbol{W}_f \), \( \boldsymbol{W}_i \), \( \boldsymbol{W}_u \), and \( \boldsymbol{W}_o \) are the weight matrices for the corresponding gates, while \( b_f \), \( b_i \), \( b_u \), and \( b_o \) are the bias terms of each gate. \( \sigma \) denotes the sigmoid activation function; \( \odot \) represents element-wise multiplication; and \( \tanh \) is the hyperbolic tangent function.   
	
	\begin{equation}
		f_t = \sigma(W_f \cdot [h_{t-1}, x_t] + b_f)
	\end{equation}
	
	\begin{equation}
		i_t = \sigma(W_i \cdot [h_{t-1}, x_t] + b_i)
	\end{equation}
	
	\begin{equation}
		{C}_t = \tanh(W_C \cdot [h_{t-1}, x_t] + b_C)
	\end{equation}
	
	\begin{equation}
		C_t = f_t \odot C_{t-1} + i_t \odot \tilde{C}_t
	\end{equation}
	
	\begin{equation}
		o_t = \sigma(W_o \cdot [h_{t-1}, x_t] + b_o)
	\end{equation}
	
	\begin{equation}
		h_t = o_t \odot \tanh(C_t)
	\end{equation}

	\section{Data Preprocessing and the proposed model}
	
	\subsection{Data Preparation and Preprocessing}
	\subsubsection{Data source}

	% % 关键：强制所有浮动体在此处“截止”，表格只能在这之后显示
	% \FloatBarrier 
	
	% % 非浮动体表格：无table环境，用\captionof指定标题
	% \centering  % 表格居中
	% \scriptsize 
	% \setlength{\tabcolsep}{3pt} 
	% \renewcommand{\arraystretch}{1} 
	% \begin{tabular}{l@{\hspace{6pt}}c@{\hspace{6pt}}c@{\hspace{6pt}}c@{\hspace{6pt}}c@{\hspace{6pt}}c@{\hspace{6pt}}c@{\hspace{6pt}}c}
		%     \toprule
		%     \textbf{Station} & \textbf{Longitude} & \textbf{Latitude} 
		%     & \multicolumn{1}{c}{\textbf{Water}} & \multicolumn{4}{c}{\textbf{SWH (m)}} \\
		%     \cmidrule(lr){4-4} \cmidrule(lr){5-8}
		%     & & & \textbf{depth (m)} & \textbf{Min} & \textbf{Max} & \textbf{Mean} & \textbf{Std} \\ 
		%     \midrule
		%     44007            & 43.525 N           & 70.141 W          & 26.5                     & 0.20         & 6.05         & 0.97          & 0.64         \\ 
		%     46087            & 48.493 N           & 124.726 W         & 260.6                    & 0.27         & 8.66         & 1.81          & 0.86         \\ 
		%     51000            & 23.535 N           & 153.781 W         & 4811.0                   & 0.89         & 8.32         & 2.31          & 0.80         \\ 
		%     \bottomrule
		% \end{tabular}
	% % 用\captionof指定表格标题（需caption宏包）
	% \label{tab:station_exact_match}

	Within the field of wave parameter analysis, satellite altimeter data, reanalysis data, and buoy data constitute three core data sources. Whilst satellite altimeter data and reanalysis data offer extensive coverage, their accuracy is relatively limited. In contrast, buoy data, as direct observational records, are regarded as the benchmark for reliability despite constraints imposed by the limited number of oceanic observation points. Given this study's focus on evaluating the performance of the WVHT forecasting model at specific stations, observed data provided by the National Oceanic and Atmospheric Administration's National Data Buoy Centre (https://www.ndbc.noaa.gov) has been employed.
	
	This study selected buoys numbered 41008 and 41047, both situated off the coast of Florida, USA, in the Atlantic Ocean waters of the southeastern North American region. 41008 being a nearshore buoy and 41047 an offshore buoy. Table 1 details the information for both buoy stations, including their respective longitude and latitude, water depth at the location, and the minimum, maximum, mean, and standard deviation of WVHT derived from statistical data. This information serves to present the location, topography, and wave characteristics of the marine sites. This study selected hourly standard meteorological data from both stations between January 2019 and December 2022, totalling approximately 66,313 records. The dataset was divided chronologically: January 2019 to December 2021 for model training, and January 2022 to December 2022 for performance evaluation. The dataset comprises 11 features: wind direction (WDIR), wind speed (WSPD), gusts (GST), significant wave height (WVHT), dominant period (DPD), average period (APD), mean wave direction (MWD), atmospheric pressure (PRES), air temperature (ATMP), water temperature (WTMP), and dew point temperature (DEWP).
	
	\begin{table}[!H]
		\caption{Statistical property of three stations.}
		\centering
		\setlength{\tabcolsep}{3pt} % 调整列间距，匹配原图紧凑感
		\renewcommand{\arraystretch}{1} % 行间距还原原图
		\scriptsize % 字体缩小，接近原图大小
		\begin{tabular}{l@{\hspace{6pt}}c@{\hspace{6pt}}c@{\hspace{6pt}}c@{\hspace{6pt}}c@{\hspace{6pt}}c@{\hspace{6pt}}c@{\hspace{6pt}}c}
			\toprule
			\textbf{Station} & \textbf{Longitude} & \textbf{Latitude} 
			& \multicolumn{1}{c}{\textbf{Water}} & \multicolumn{4}{c}{\textbf{SWH (m)}} \\
			\cmidrule(lr){4-4} \cmidrule(lr){5-8}
			& & & \textbf{depth (m)} & \textbf{Min} & \textbf{Max} & \textbf{Mean} & \textbf{Std} \\ 
			\midrule
			41008            & 31.400 N           &  80.866 W         & 16                       & 0.20         & 4.54         & 1.13          & 0.49         \\ 
			41047            & 27.557 N           &  71.480 W         & 5328                     & 0.49         & 9.34         & 1.63          & 0.76         \\ 
			\bottomrule
		\end{tabular}
		
		\label{tab:station_exact_match}
	\end{table}

	\subsubsection{Filling missing values}
	Marine data collection often suffers from data loss due to equipment failures, harsh environments, or transmission issues, and most machine learning models cannot directly process datasets with missing values, missing value imputation is a critical step in data preprocessing. Therefore, this study employs weighted linear interpolation to sequentially repair missing values in 11 features, ensuring the temporal continuity of the data. This method is simple and efficient, suitable for the characteristics of marine data. It dynamically assigns weights based on the temporal distance between missing values and neighbouring valid data points, with closer observation points receiving higher weights, thereby improving interpolation accuracy. Its mathematical expression is shown in as in formula (13).

	\begin{equation}
		x_{\text{filled}} = \left( \frac{t_{\text{next}} - t_{\text{miss}}}{t_{\text{next}} - t_{\text{prev}}} \right) x_{\text{prev}} + \left( \frac{t_{\text{miss}} - t_{\text{prev}}}{t_{\text{next}} - t_{\text{prev}}} \right) x_{\text{next}}
	\end{equation}
	where \( x_{\text{filled}} \) is the interpolated result of the missing value; \( x_{\text{prev}} \) and \( x_{\text{next}} \) are the first valid observation data before and after the missing value, respectively; \( t_{\text{prev}} \), \( t_{\text{miss}} \), and \( t_{\text{next}} \) correspond to the timestamps or sequence position indices of \( x_{\text{prev}} \), the missing value, and \( x_{\text{next}} \), respectively.

	\subsubsection{Preprocessing angle feature}
	
	\begin{figure}[h]  % [H] = Here，强制在此处插入，不浮动
		\centering
		% 建议保留一个includegraphics即可（避免重复插入）
		\includegraphics[width=1\linewidth]{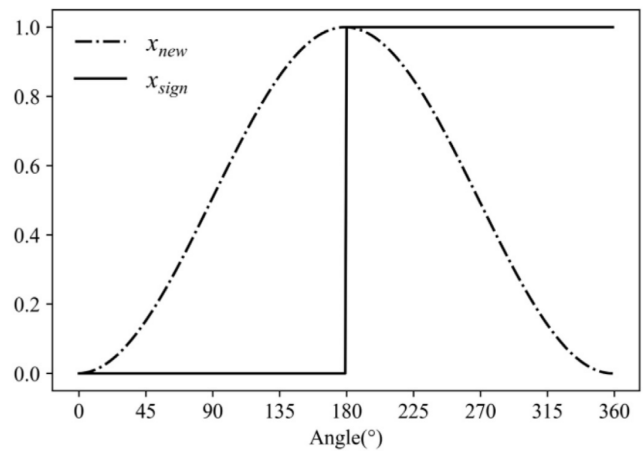}  
		\vspace{-17pt} 
		\caption{Symbols of Transformed Values and Angular Features}
		% 图标题 
		\label{fig:feature_extraction}  % 标签（引用时用\ref{fig:feature_extraction}）
	\end{figure}

	Angular features such as WDIR, MWD, ATMP, WTMP, and DEWP differ fundamentally from other features such as WVHT due to their periodic nature. For example, 1° and 360° are almost equivalent in physical terms, but there is a significant numerical difference. This characteristic makes it difficult for traditional numerical processing methods to effectively handle angular features, especially in machine learning models, which may lead to low learning efficiency or distorted prediction results. To address this, this study proposes a reasonable and reversible angular feature preprocessing method, converting the original angle value \( x \) into a tuple (\( x_{\text{new}}, x_{\text{sign}} \)) to retain periodic physical meaning and improve model training effectiveness.

	The specific conversion formulas are (14) and (15).
	\begin{equation}
		x_{\text{new}} = -0.5 \cos\left( \frac{2\pi x}{T} \right) + 0.5
	\end{equation}
	\begin{equation}
		x_{\text{sign}} = \begin{cases} 
			1, & x > 0.5T \\
			0, & x \leq 0.5T 
		\end{cases}
	\end{equation}
	where \( T \) is the period (usually 360°), \( x_{\text{new}} \) is the normalized angle value, and \( x_{\text{sign}} \) is the converted sign identifier, distinguishing the relative position of the angle within the period. Through periodic mapping of the cosine function, angle values are normalized to the [0, 1] interval, ensuring that angles with close numerical values (for example, 1° and 360°) have similar \( x_{\text{new}} \) after conversion, effectively retaining periodic characteristics. This method not only promotes model convergence but also avoids learning difficulties caused by numerical differences. The mapping relationship of the converted features is shown in Figure 1.

	\subsubsection{Min-max normalization}
	
	\begin{figure*}[b]  % [H] = Here，强制在此处插入，不浮动
		\centering
		% 建议保留一个includegraphics即可（避免重复插入）
		\includegraphics[width=1\textwidth]{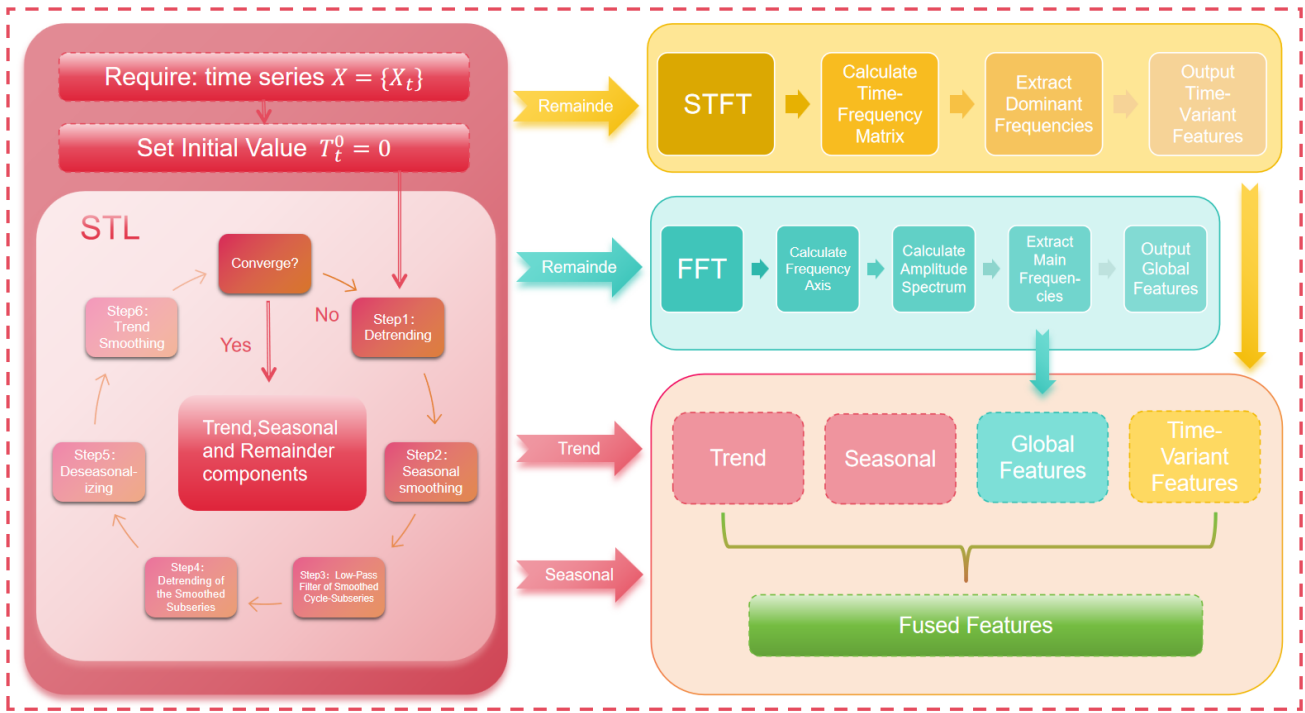}  
		\vspace{-17pt} 
		\caption{Feature extraction module}  % 图标题
		\label{fig:feature_extraction}  % 标签（引用时用\ref{fig:feature_extraction}）
	\end{figure*}
	
	The characteristic dimensions of ocean and meteorological time series differ significantly, and direct use of raw data may lead to unstable model training or slow convergence. To eliminate dimensional differences, this study uses the minimum-maximum normalisation method to map the 11 feature values processed in 3.1.2 and 3.1.3 to the [0, 1] interval, as in formula (16).

	\begin{equation}
		x_{\text{scaled}} = \frac{x - x_{\text{min}}}{x_{\text{max}} - x_{\text{min}}}
	\end{equation}
	where \( x \) is the original feature value, \( x_{\text{min}} \) and \( x_{\text{max}} \) are the minimum and maximum values of the feature in the dataset, respectively, and \( x_{\text{scaled}} \) is the normalized feature value, with a range standardized to [0, 1]. Normalization unifies feature scales, improving the stability and convergence speed of model training.
	
	\subsection{Feature extraction module (STL-FFT-STFT)}

	Feature extraction is a critical step in complex time series analysis. This study proposes an integrated feature extraction module that combines STL, FFT, and STFT to perform multidimensional feature extraction and analysis of complex signals from three levels: time domain decomposition, global frequency domain analysis, and time-frequency localisation analysis. Eleven data features are extracted, including trend features, seasonal features, global spectral features, and time-frequency spectral features, as shown in Figure 2.
	
	First, STL decomposes the 11 normalised features one by one to extract the trend component \( T(t) \), seasonal component \( S(t) \), and residual component \( R(t) \), as in formula (17). By removing the deterministic components of trend and seasonality, the residuals are closer to high-frequency noise and extreme wave height sudden change signals, thereby providing high-quality input data for feature extraction. The trend component represents long-term evolutionary patterns, while the seasonal component captures periodic fluctuations. STL is based on Loess and is robust to outliers, enabling it to effectively separate the structured features of time series, thereby providing high-quality input for subsequent feature extraction and model training. Taking the STL decomposition of WVHT for the entire year of 2019 at site 41008 as an example,the decomposition results are shown in Figure 3.
	\begin{equation}
		x_{\text{scaled}} = T(t) + S(t) + R(t)
	\end{equation}
	
	\begin{figure}[h]  % [H] = Here，强制在此处插入，不浮动
		\centering
		% 建议保留一个includegraphics即可（避免重复插入）
		\includegraphics[width=1\linewidth]{stl.pdf}  
		\vspace{-17pt} 
		\caption{WVHT 2019 STL Decomposition Plot}  % 图标题
		\label{fig:feature_extraction}  % 标签（引用时用\ref{fig:feature_extraction}）
	\end{figure}
	
	Next, to further explore the frequency characteristics of the residual component \( R_t \) and address the issue of reduced prediction accuracy caused by noise points generated in abnormal weather, the FFT, as a classic frequency - domain analysis tool, sequentially performs global frequency - domain transformation on the residual components \( R_t \) obtained from the decomposition of the 11 features. This transforms them into frequency - domain representations \( y_{f} \) and acquires the overall spectral characteristics. First, remove the NaN values in the residuals, determine the length of valid data \( N=len (residual) \), and generate the frequency axis \( x_{f} \) according to the sampling rate. By calculating the maximum - minimum normalization of the amplitude spectrum, the significant frequency components with amplitudes exceeding 20\% the maximum value are selected, and the main periods are extracted as the global frequency features. Taking the FFT decomposition of WVHT throughout 2019 at site 41008 as an example, the decomposition results are shown in Figure 4. However, FFT cannot accurately capture the time - varying characteristics of signals, so it is necessary to combine STFT for supplementary analysis.  
	
	% 假设图 3.4.2 的插入代码如下，需根据实际文件路径和格式调整
	\begin{figure}[h]  % [H] = Here，强制在此处插入，不浮动
		\centering
		% 建议保留一个includegraphics即可（避免重复插入）
		\includegraphics[width=1\linewidth]{fft.pdf}  
		\vspace{-17pt} 
		\caption{WVHT 2019 FFT Decomposition Plot}  % 图标题
		\label{fig:feature_extraction}  % 标签（引用时用\ref{fig:feature_extraction}）
	\end{figure}

	Finally, to capture the time-varying frequency characteristics of residuals and track the temporal nodes and trend shifts arising from data discontinuities in non-linear dynamic processes, the STFT employs a sliding window to segment the residual signal into multiple segments. The window length is set via nperseg=128, meaning each sliding window encompasses 128 data points. Furthermore, noverlap=64 configures the window overlap length to 64 data points, resulting in a 50\% overlap ratio. This approach helps mitigate spectral leakage between adjacent windows, yielding smoother time-frequency analysis results. Concurrently, as no explicit window function type is specified, the default Hanning window is employed. This windowing function achieves a favourable balance between temporal and frequency resolution. It performs FFT on each segment to generate a time-frequency matrix \( Z_{xx} \). Then, it calculates the amplitude spectrum to form a time-frequency heatmap, and extracts the dominant frequency sequence of each time window. This reveals the dynamic changes of frequency components and makes up for the deficiency of FFT in time-varying analysis. STFT is particularly suitable for the analysis of non-stationary time series. Taking the STFT decomposition of WVHT throughout 2019 at site 41008 as an example, the decomposition result is shown in Figure 5.
	
	\begin{figure}[h]  % [H] = Here，强制在此处插入，不浮动
		\centering
		% 建议保留一个includegraphics即可（避免重复插入）
		\includegraphics[width=1\linewidth]{stft.pdf}  
		\vspace{-17pt} 
		\caption{WVHT 2019 STFT Decomposition Plot}  % 图标题
		\vspace{-17pt} 
		\label{fig:feature_extraction}  % 标签（引用时用\ref{fig:feature_extraction}）
	\end{figure}
	
	\subsection{Network module (TCN-LSTM)}

	The network module is the core of the model's prediction capabilities, effectively modelling the complex interactions within time series data by combining TCN and LSTM (\cite{Klemm2023PredictingFutureWaveHeights}). In this study, the input features of the network module include trend features, seasonal features, global spectral features, and time-frequency spectral features. Through the synergistic effect of TCN-LSTM, the model can effectively learn the dynamic patterns and interactions of time series, providing support for high-precision predictions. The overall architecture of the network module is shown in Figure 6.
	
	\begin{figure*}[B]  % [H] = Here，强制在此处插入，不浮动
		\centering
		\includegraphics[width=1\textwidth]{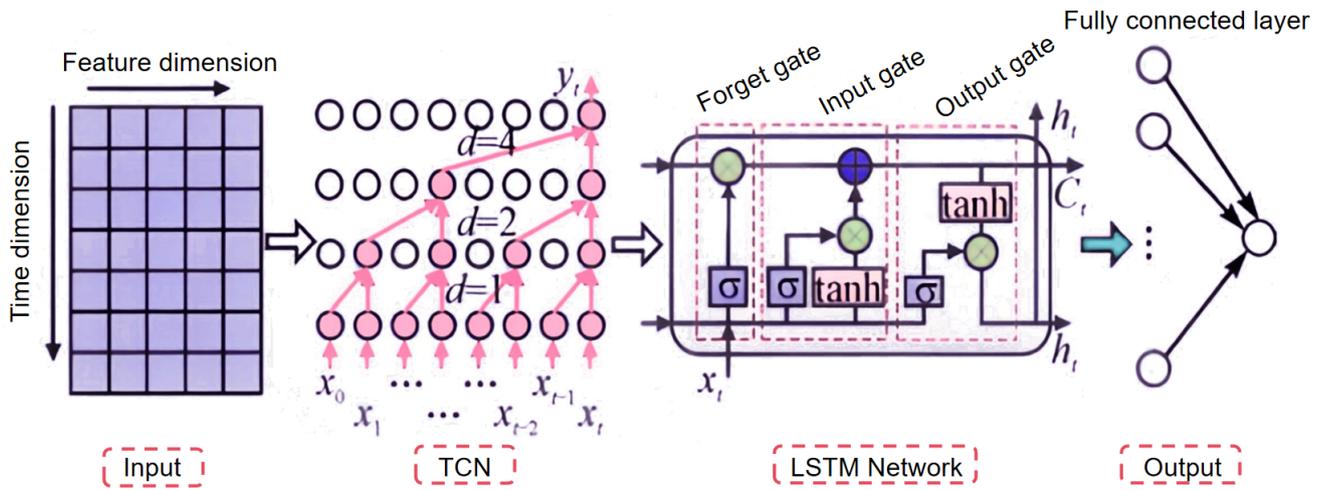}  
		\vspace{-17pt} 
		\caption{Network module}  % 图标题
		\label{fig:feature_extraction}  % 标签（引用时用\ref{fig:feature_extraction}）
	\end{figure*}

	TCN leverages its expansion causal convolution structure to achieve exceptional feature extraction capabilities, enabling the fusion of multi-source features to generate high-dimensional abstract representations and enhance the depth of feature information extraction. The TCN architecture consists of multiple TCN blocks, each containing two one-dimensional expansion convolution layers (Conv1D) with expansion rates of 1, 2, and 4, capturing features across different temporal scales. Each layer is followed by Batch Normalisation, ReLU activation functions, and Dropout layers (probability p=0.2) to accelerate training and prevent overfitting. TCN blocks introduce residual connections, which adjust the input dimension via 1x1 convolutions before adding them to the output, mitigating the vanishing gradient problem. In this study, the 11 features involved exhibit complex interrelationships. As visually demonstrated by the correlation heatmap in Figure 7, taking the 41008 site on the left as an example, the correlation coefficient between ATMP\_new and WTMP\_new reaches 0.95, and that between WSPD\_scaled and GST\_scaled is 0.99. The associations among various features are rich and diverse. Temporal Convolutional Network (TCN) leverages such feature correlations to fuse trend, seasonality, global spectrum, and time-varying features. By virtue of its expanded causal convolution structure, it fully explores the associations among multi-source features, generating high-dimensional abstract representations. These representations serve as high-quality inputs for subsequent Long Short-Term Memory (LSTM) prediction, enabling the model to better utilize the correlation information among features and enhance prediction performance.
	
	\begin{figure*}[b]  % [H] = Here，强制在此处插入，不浮动
		\centering
		\includegraphics[width=1\textwidth]{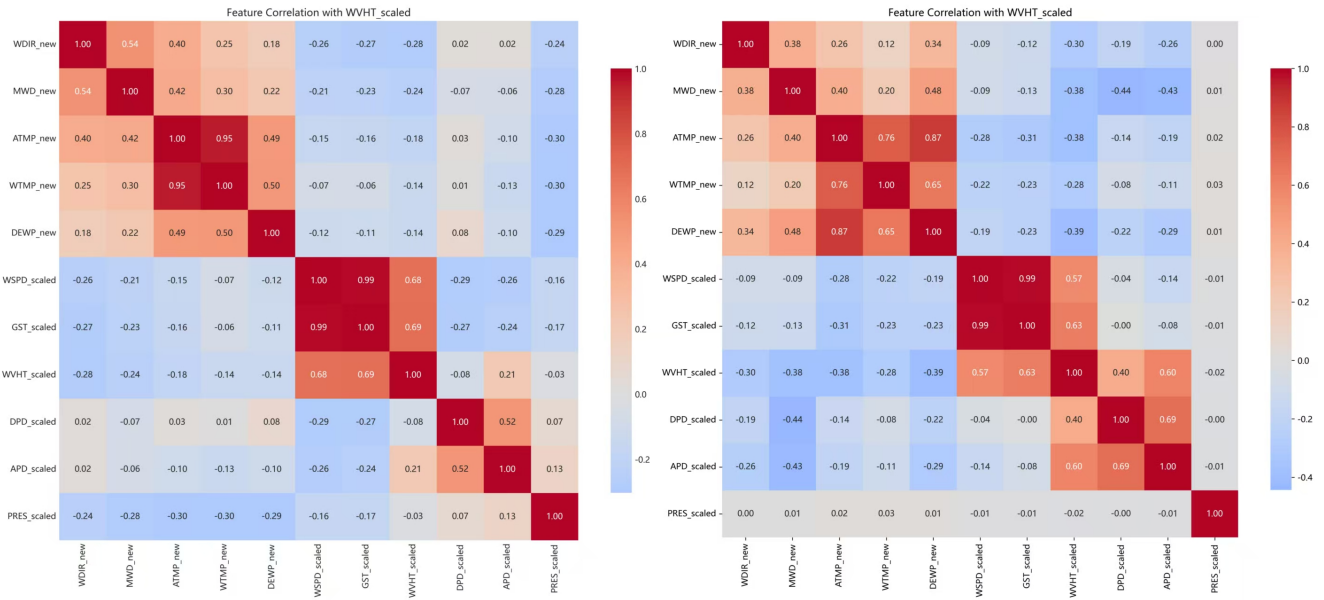}  
		\vspace{-17pt} 
		\caption{Correlation Analysis Diagram(Left: Site 41008; Right: Site 41047 )}  % 图标题
		\label{fig:feature_extraction}  % 标签（引用时用\ref{fig:feature_extraction}）
	\end{figure*}

	LSTM receives the high-dimensional features extracted by TCN and further models the long-term dependencies of the time series to generate accurate prediction results. LSTM captures the dynamic patterns in the sequence through a gating mechanism and ultimately outputs the prediction results through a fully connected layer. The model is compiled using the Adam optimiser (learning rate 0.001) with mean squared error (MSE) as the loss function, and the optimisation objective is to minimise the error between the predicted values and the actual values(\cite{Kingma2014Adam}). Model performance is evaluated using the mean absolute error (MAE) as the assessment metric. By combining TCN with LSTM, TCN first efficiently extracts and integrates the input features, then feeds the processed features into LSTM for time series prediction, thereby enhancing the model's efficiency and accuracy in modelling complex time series.

	\section{Experiments}
	
	\subsection{Experimental platform}
	To ensure fairness in performance comparisons and reproducibility of experimental results, all experiments were conducted on a unified instance of the AutoDL platform (https://www.autodl.com). This platform is equipped with four NVIDIA RTX 4090 GPUs running the Ubuntu operating system, providing a stable and efficient computational environment. Experiments were developed using Python 3.9, leveraging open-source toolkits such as TensorFlow, Pandas, and NumPy to implement the algorithmic framework. These tools provided robust support for data processing, model construction, and performance evaluation, ensuring an efficient experimental process and reliable results.

	\subsection{Metrics}
	To comprehensively evaluate the prediction performance of the model, this study adopts the following five commonly-used indicators: Root Mean Square Error (RMSE), Mean Absolute Error (MAE), Symmetric Mean Absolute Percentage Error (SMAPE), R-square ($R^2$), and Correlation Coefficient (CC). These indicators quantify the deviation and similarity between the predicted values and the true values from different perspectives. Formula (18) is used to reflect the overall deviation. The smaller the RMSE, the smaller the overall prediction deviation of the model, especially the better the control over ``samples with large errors''. Formula (19) is used to measure the average absolute deviation between the predicted values and the true values. The smaller the value of MAE, the smaller the deviation between the predicted values and the true values, and the better the prediction performance of the model. Formula (20) is a symmetric percentage error indicator, which avoids the calculation abnormality of the traditional MAPE due to ``too small true values''. The smaller the SMAPE, the better the effect. Formula (21) explains the proportion of the variation of the dependent variable to the total variation of the dependent variable. The closer $R^2$ is to 1, the smaller the proportion of the model error to the total difference, and the better the fitting effect. Formula (22) measures the similarity between the predicted value sequence and the true value sequence. The value range of CC is [-1, 1]. The closer the value is to 1, the higher the sequence similarity between the predicted values and the true values, and the better the prediction performance of the model. Among them, $\hat{y}_i$ is the predicted value; $y_i$ is the true value; $\bar{y}$ is the mean value; and $n$ is the number of samples.

	\begin{equation}
		\text{RMSE} = \sqrt{\frac{1}{n} \sum_{i=1}^{n} \left( \hat{y}_i - y_i \right)^2}
	\end{equation}
	
	\begin{equation}
		\text{MAE} = \frac{1}{n} \sum_{i=1}^{n} \left| \hat{y}_i - y_i \right|
	\end{equation}
	
	\begin{equation}
		\text{SMAPE} = \frac{100\%}{n} \sum_{i=1}^{n} \frac{\left| \hat{y}_i - y_i \right|}{\left( \left| \hat{y}_i \right| + \left| y_i \right| \right) / 2}
	\end{equation}
	
	\begin{equation}
		R^2 = 1 - \frac{\sum_{i=1}^{n} \left( y_i - \hat{y}_i \right)^2}{\sum_{i=1}^{n} \left( y_i - \bar{y} \right)^2}
	\end{equation}
	
	\begin{equation}
		\text{CC} = \frac{\sum_{i=1}^{n} \left( \hat{y}_i - \bar{y} \right) \left( y_i - \bar{y} \right)}{\sqrt{\sum_{i=1}^{n} \left( \hat{y}_i - \bar{y} \right)^2 \sum_{i=1}^{n} \left( y_i - \bar{y} \right)^2}}
	\end{equation}

	\subsection{Comparative experiment}
	
	To achieve precise WVHT forecasting, this study addresses two key challenges—capturing extreme wave heights and suppressing high-frequency noise—by proposing a hybrid STL-FFT-STFT-TCN-LSTM model. Through multi-scale decomposition via STL, FFT, and STFT combined with deep temporal modelling using TCN-LSTM, this approach significantly enhances predictive capability under extreme conditions. Comparative experiments selected ANN, LSTM, EMD-LSTM, TCN-LSTM, and STL-CNN-PE as representative baselines. Validation was conducted across two distinct marine environments: the nearshore site 41008 (water depth 16m) and the offshore site 41047 (water depth 5328m). Results were systematically analysed through scatter plots, time series, and waterfall plots.

	\subsubsection{Scatter plot}

	\begin{figure*}[b]
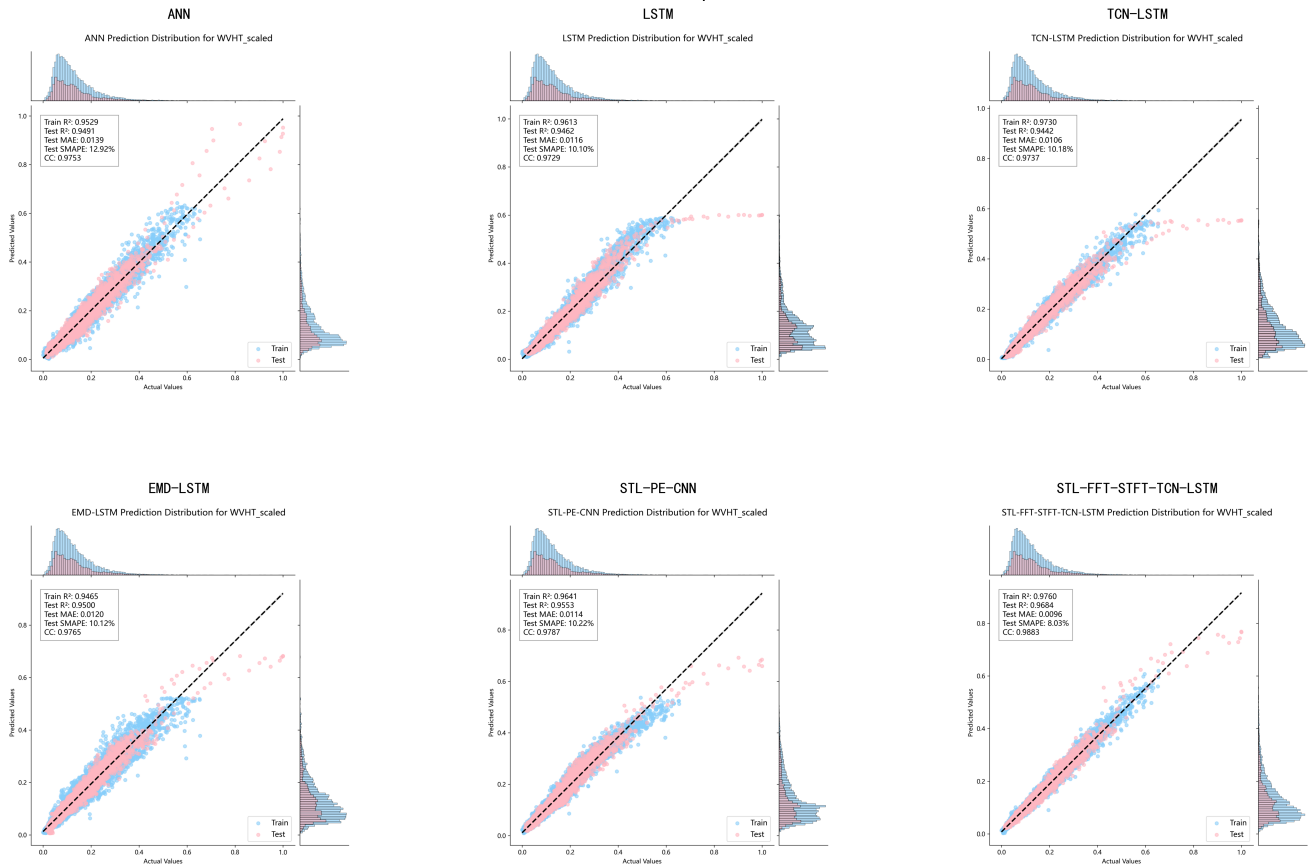
  % [H] = Here，强制在此处插入，不浮动
		\centering
		\includegraphics[width=1\textwidth]{bluepink41008.pdf}  
		\includegraphics[width=1\textwidth]{bluepink41047.pdf}  
		\vspace{-17pt} 
		\caption{Scatter plot(Top: Site 41008; Bottom: Site 41047)}  % 图标题
		\label{fig:feature_extraction}  % 标签（引用时用\ref{fig:feature_extraction}）
	\end{figure*}
	
	The results of the scatter plots reveal the performance differences of various models in different marine environments, as shown in Figure 8.

	At the nearshore site 41008, the baseline models ANN, LSTM, EMD-LSTM, TCN-LSTM, and STL-CNN-PE all exhibit significant deviations in scenarios of extreme wave heights and high-frequency noise. For example, in the interval where the actual values are greater than 0.8, the prediction results of ANN deviate significantly from the diagonal, with a Test MAE reaching 0.0224. In contrast, the MAE of STL-FFT-STFT-TCN-LSTM is only 0.0133, representing a 40.5\% reduction in error. The overall performance of LSTM is better than that of ANN, with a Train \( R^2 \) of 0.9680 and a Test \( R^2 \) of 0.9519. However, there is still obvious dispersion in the prediction of extreme value intervals close to 1.0, and the SMAPE of LSTM is 11.82\%, indicating insufficient noise suppression. In comparison, the prediction results of STL-FFT-STFT-TCN-LSTM in the high-value region greater than 0.8 almost coincide with the true values. Its Train \( R^2 \) is 0.9837, Test \( R^2 \) is 0.9789, and CC is 0.9911, demonstrating significant adaptability to the complex nearshore environment.

	At the offshore site 41047, influenced by ocean currents and deep - layer dynamics, the fluctuations are more complex. The superposition of extreme wave heights and noise poses greater challenges to the models. STL - CNN - PE shows a large degree of dispersion in predictions when the actual values are greater than 0.8. Its Test \( R^2 \) is 0.9553, and the MAE is 0.0114, with obvious errors in extreme values. Although TCN - LSTM has a good overall fit, it has more outliers due to noise interference, and its SMAPE is 10.18\%. In comparison, the STL - FFT - STFT - TCN - LSTM model shows a significant improvement in the extreme value region. Its Test \( R^2 \) is 0.9684, which is 1.31\% higher than that of STL - CNN - PE. The Test MAE is 0.0096, and the SMAPE is 8.03\%, demonstrating robustness in the complex offshore scenarios.
	
	Comprehensive analysis indicates that Station 41008 experiences frequent noise influenced by nearshore currents, yet exhibits distinct fluctuation patterns. The STL-FFT-STFT-TCN-LSTM approach decomposes trends and seasonal components via STL, effectively separates high-frequency noise using FFT and STFT, and achieves precise fitting through TCN-LSTM. This reduces extreme wave height prediction errors by 40.5\%. At Station 41047, the model captures both long- and short-period fluctuations—such as ocean currents and wind waves—through multi-scale decomposition, reducing errors by 15.8\%–20.3\%. Overall, the STL-FFT-STFT-TCN-LSTM model reduced MAE by 15.8\%-40.5\% and SMAPE by 8.3\%-20.3\% for extreme wave heights exceeding 0.8 metres. This fully validates its high accuracy and noise suppression capabilities across diverse marine conditions, providing reliable support for wave energy development.

	\subsubsection{Time series analysis diagram}
	
	Time series analysis further reveals the temporal prediction performance of the models, as shown in Figure 9.  
	
	At the nearshore site 41008, there are significant extreme peaks in the time steps of 4000 - 8000, such as at 6000 and 7500. ANN only predicts 70\% of the true value at the 6000 peak, and LSTM has a deviation of more than 20\% at the 7500 peak. In comparison, the prediction curve of the STL-FFT-STFT-TCN-LSTM almost coincides with the true value. The deviation at the 6000 peak is less than 5\%, and the deviation at the 7500 peak is less than 3\%, indicating its advantage in capturing extreme wave heights. In the high-frequency noise scenario of the 0 - 2000 time steps, the curve of EMD-LSTM oscillates violently, with a deviation exceeding 15\% in the 500 - 1000 interval; TCN-LSTM is over-smoothed, and small peaks at 1500 are eliminated. The STL-FFT-STFT-TCN-LSTM model separates high-frequency noise through STL decomposition, FFT, and STFT, and then focuses on core fluctuations by TCN-LSTM, achieving a deviation of less than 8\%, which is significantly better than other models.  
	
	At the offshore site 41047, the extreme peak in the time steps of 5000 - 6000 reflects sudden events, such as mesoscale eddies. EMD-LSTM and STL-CNN-PE failed to accurately capture the peak. The predicted amplitude of the former was only 60\% of the true value. In contrast, the STL-FFT-STFT-TCN-LSTM model accurately identified the peak time and amplitude, with a deviation of less than 5\%, verifying its robustness in offshore conditions. In the time steps of 0 - 3000, the true values showed a coupling of long-term cycles and high-frequency fluctuations. ANN misaligned in frequency identification, and LSTM lost high-frequency details, with a deviation exceeding 20\% at 2500. The STL-FFT-STFT-TCN-LSTM model effectively distinguished different frequency components through multi-scale decomposition and used TCN-LSTM to accurately model complex coupling scenarios, with an overall deviation of less than 10\%.  
	
	The cross-site comparison shows that site 41008 has regular peaks but dense noise, while site 41047 has abrupt peaks and is interfered by complex currents. The STL-FFT-STFT-TCN-LSTM model achieved a peak amplitude deviation of less than 5\%, a temporal deviation of less than 10 steps, and a noise-induced deviation of less than 10\% in both environments, demonstrating strong adaptability across sea areas.

	\begin{figure*}[b]  % [H] = Here，强制在此处插入，不浮动
		\centering
		\includegraphics[width=1\textwidth]{yuce.pdf}  
		\vspace{-17pt} 
		\caption{Comparison Chart of Actual and Forecast Values (Top: Site 41008; Bottom: Site 41047)}  % 图标题
		\label{fig:feature_extraction}  % 标签（引用时用\ref{fig:feature_extraction}）
	\end{figure*}

	\subsubsection{Waterfall plot}
	
	The waterfall plot intuitively reveals the performance differences among various models from the three - dimensional perspective of metric - model - value, as shown in Figure 10.  
	
	At site 41008, the purple surface of STL-FFT-STFT-TCN-LSTM shows that the MAE is 0.0133 and the SMAPE is 0.8231, which is significantly better than that of ANN (with an MAE of 0.0224 and an SMAPE of 0.1551). Meanwhile, the CC of this model is 0.99 and \( R^2 \) is 0.98, which are also higher than the CC (0.97) and \( R^2 \) (0.96) of TCN-LSTM, demonstrating stronger fitting performance.  
	
	At site 41047, although the overall index performance is slightly inferior to that in the near - shore area, STL-FFT-STFT-TCN-LSTM still maintains advantages. The MAE is 0.0096, the SMAPE is 0.0803, and the CC is 0.9883 and \( R^2 \) is 0.9684, all exceeding the CC (0.9737) and \( R^2 \) (0.9442) of TCN-LSTM.  
	
	The cross - sea - area comparison results show that at site 41008, this model achieves a 10\% - 30\% reduction in MAE/SMAPE and a 5\% - 10\% improvement in CC/\( R^2 \). At site 41047, the error is reduced by 20\% - 40\% and the performance is improved by 8\% - 15\%. It effectively overcomes the core difficulties of extreme wave height prediction and noise suppression, providing a robust cross - sea - area solution for WVHT prediction.

	\begin{figure*}[b]  % [H] = Here，强制在此处插入，不浮动
		\centering
		\includegraphics[width=1\textwidth]{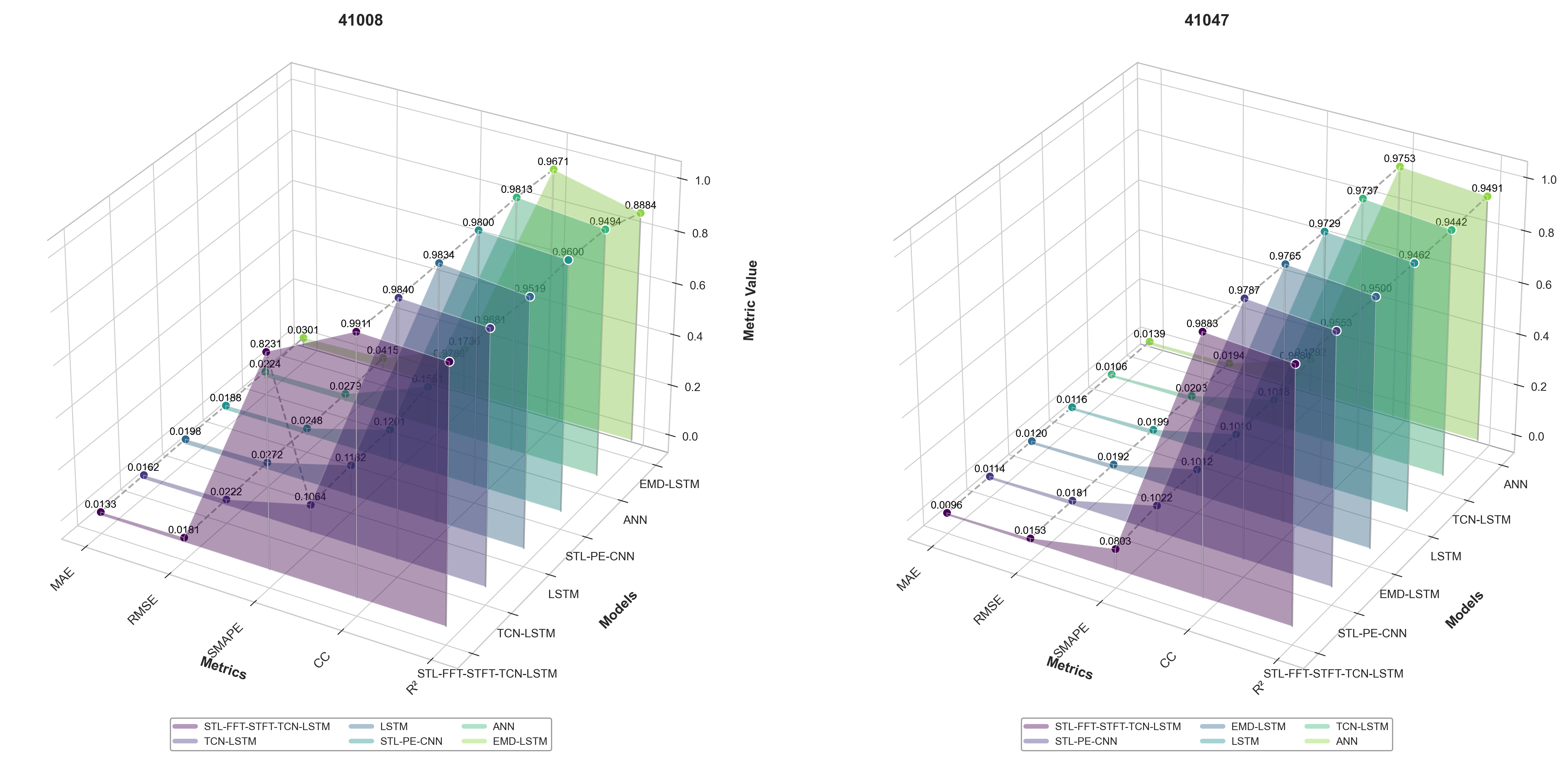}  
		\vspace{-17pt} 
		\caption{Waterfall plot(Left: Site 41008; Right: Site 41047 )}  % 图标题
		\label{fig:feature_extraction}  % 标签（引用时用\ref{fig:feature_extraction}）
	\end{figure*}

	\subsection{Ablation study}
	
	To verify the necessity of the core components in the STL-FFT-STFT-TCN-LSTM model and the superiority of its multi-scale feature fusion, this study designs four sets of ablation experiments at the 41008 site, as shown in Table 2. Taking the complete model including STL decomposition, FFT, STFT, and TCN-LSTM as the baseline, an $\Delta$STL model with STL decomposition removed, an $\Delta$FFT model with FFT removed, an $\Delta$STFT model with STFT removed, and an $\Delta$Both model with all frequency-domain features removed while only retaining STL decomposition are constructed respectively. Through radar charts, scatter plots, box plots, and bar charts, the indispensable role of each component is systematically analyzed based on measured data.

	\begin{table}[htbp]
		\centering
		\setlength{\tabcolsep}{3pt} % 调整列间距，匹配原图紧凑感
		\caption{Model Performance Comparison in Ablation Study}
		\label{tab:ablation_study}
		\begin{tabular}{l|cccccc}
			\hline
			\textbf{Model} & \textbf{MAE} & \textbf{RMSE} & \textbf{SMAPE (\%)} & \textbf{CC} & \textbf{R²} \\
			\hline
			Baseline & 0.0124 & 0.0172  & 7.6453 & 0.9906 & 0.9809 \\
			$\Delta$ STL & 0.0178 & 0.0243  & 10.7315 & 0.9848 & 0.9616 \\
			$\Delta$ FFT & 0.0228 & 0.0270 & 17.0093 & 0.9911 & 0.9528 \\
			$\Delta$ STFT & 0.0149 & 0.0201  & 9.1801 & 0.9871 & 0.9727 \\
			$\Delta$ Both & 0.0169 & 0.0227  & 9.5126 & 0.9860 & 0.9652 \\
			\hline
		\end{tabular}
	\end{table}

	As shown in Figure 11, the radar chart, with normalized RMSE, MAE, SMAPE, CC, and \( R^2 \) as dimensions, displays the comprehensive performance of each model. At site 41008, the baseline model shows the largest coverage area and the optimal indicators: RMSE = 0.0172, MAE = 0.0124, SMAPE = 7.6453\%, CC = 0.9906, and \( R^2 = 0.9809 \), reflecting the balanced superiority of multi - scale fusion in error control and fitting degree. The area of the \( \Delta \text{STL} \) model shrinks significantly, with RMSE increasing to 0.0243, MAE increasing to 0.0178, CC decreasing to 0.9848, and \( R^2 \) decreasing to 0.9616, highlighting the key role of STL decomposition in capturing structural trends and periodic fluctuations. The error of the \( \Delta \text{FFT} \) model intensifies, with MAE increasing to 0.0228 and SMAPE increasing to 17.0093\%, indicating the irreplaceability of FFT for the global frequency distribution. The fitting degree of the \( \Delta \text{STFT} \) model decreases, with \( R^2 \) decreasing to 0.9727 and CC decreasing to 0.9871, verifying the necessity of STFT in dynamic frequency capture. The \( \Delta \text{Both} \) model has the smallest area, with SMAPE reaching 9.5126\% (about 1.24 times that of the baseline) and \( R^2 \) decreasing to 0.9652, confirming the crucial importance of the synergistic fusion of frequency - domain features and STL.

	\begin{figure}[h]  % [H] = Here，强制在此处插入，不浮动
		\centering
		% 建议保留一个includegraphics即可（避免重复插入）
		\includegraphics[width=1\linewidth]{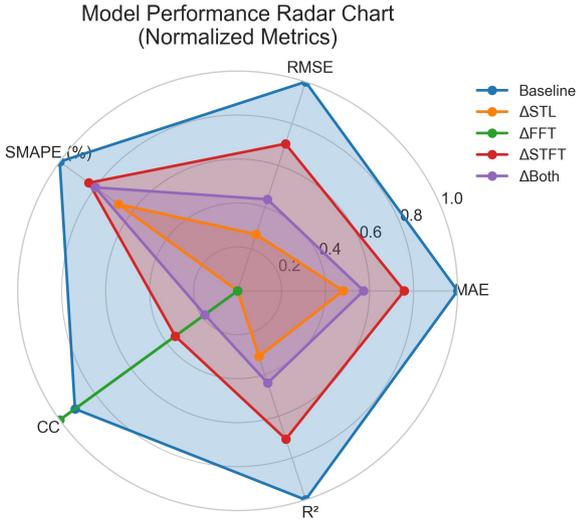}  
		\vspace{-17pt} 
		\caption{Radar chart}  % 图标题
		\label{fig:feature_extraction}  % 标签（引用时用\ref{fig:feature_extraction}）
	\end{figure}
	
	As shown in Figure 12, the scatter plot indicates that the closer the data points are to the diagonal line $y = x$, the higher the degree of coincidence between the predicted values and the true values. At site 41008, the baseline model shows a characteristic of dense adherence to the line. In the extreme wave height region where the true value is greater than 0.8, the deviation is less than 5\%, and in the high - frequency noise region of 0 - 2000 steps, the dispersion degree is less than 8\%, reflecting the dual control ability of multi - scale fusion over extreme values and noise. The scatter points of the $\Delta$STL model are obviously discrete. In the extreme wave height region where the true value is 1.0, the deviation from the diagonal line is 15\% - 20\%, and the noise region shows a radial diffusion, lacking trend constraints. For the $\Delta$FFT model, in the global frequency - sensitive region where the true value is 0.5 - 0.7, the deviation is 10\% - 12\%, while it is less than 5\% for the baseline model, indicating that the lack of FFT leads to the macroscopic periodic characteristics. The deviation of the $\Delta$STFT model in the time - varying fluctuation region of 0 - 1000 steps reaches 12\%, which is 1.5 times that of the baseline, confirming the indispensability of STFT for dynamic frequencies. The $\Delta$Both model has the most outliers, and the deviation in the extreme value and noise regions exceeds 20\%, emphasizing the necessity of complete multi - scale feature fusion.

	\begin{figure}[h]  % [H] = Here，强制在此处插入，不浮动
		\centering
		% 建议保留一个includegraphics即可（避免重复插入）
		\includegraphics[width=1\linewidth]{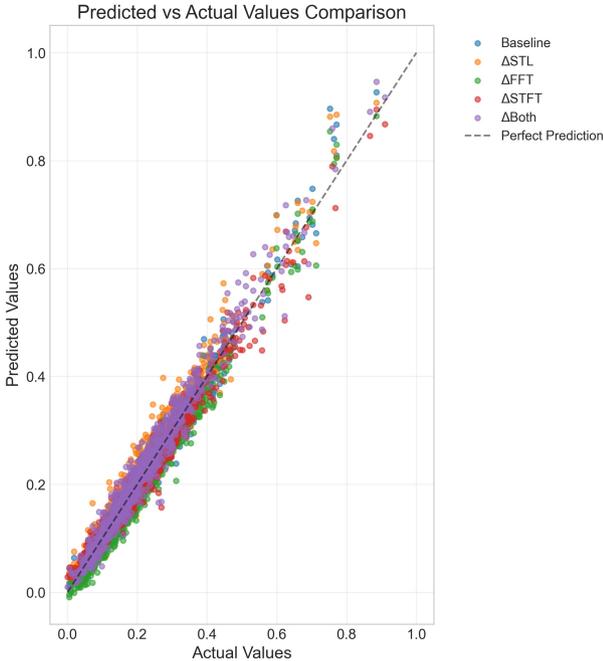}  
		\vspace{-17pt} 
		\caption{Scatter plot}  % 图标题
		\label{fig:feature_extraction}  % 标签（引用时用\ref{fig:feature_extraction}）
	\end{figure}

	As shown in Figure 13, the box plot and bar chart are presented. The box plot evaluates the model stability from the perspectives of the concentration, dispersion, and outliers of the error distribution. The baseline model has the narrowest box. The interquartile range (IQR) is approximately 0.008, the median is close to 0, and the proportion of outliers is less than 3\%, indicating that the error distribution is concentrated, the fluctuation is small, and the stability is excellent. The width of the box of the $\Delta$STL model is about twice that of the baseline. Its IQR is approximately 0.016, the median deviates from 0 by 0.012, and the proportion of outliers increases to 15\%, indicating that the error loses constraints after the STL is missing, and the fluctuation intensifies. The IQR of the $\Delta$FFT model is 0.013, and the proportion of outliers is 8\%, concentrated in the frequency mutation region, reflecting the importance of global frequency - domain features to stability. The IQR of the $\Delta$STFT model is 0.012, and most outliers are distributed in the high - frequency band, verifying the role of time - varying frequency - domain features in local error control. The box of the $\Delta$Both model is the widest, with an IQR of approximately 0.021 and an outlier proportion of 22\%, showing an imbalance between trend and dynamics and the worst stability. The bar chart quantifies the differences of each model in key indicators. At site 41008, the MAE of the baseline model is reduced by approximately 30.3\% compared to the $\Delta$STL model, by approximately 45.6\% compared to the $\Delta$FFT model, by approximately 16.8\% compared to the $\Delta$STFT model, and by approximately 26.6\% compared to the $\Delta$Both model; $R^2$ is increased by approximately 2.0\% compared to the $\Delta$STL model, by approximately 2.9\% compared to the $\Delta$FFT model, by approximately 0.8\% compared to the $\Delta$STFT model, and by approximately 1.6\% compared to the $\Delta$Both model. These data prove that STL decomposition, FFT, and STFT are each irreplaceable: STL provides a structural basis, FFT captures global frequencies, STFT adapts to time - varying features, and when fused with TCN - LSTM, forms a multi - scale system of trend - global frequency - time - varying frequency.

	\begin{figure*}[b]  % [H] = Here，强制在此处插入，不浮动
		\centering
		\includegraphics[width=1\textwidth]{ronghe3.pdf}  
		\vspace{-17pt} 
		\caption{Comparison Chart of Actual and Forecast Values (Top: Site 41008; Bottom: Site 41047)}  % 图标题
		\label{fig:feature_extraction}  % 标签（引用时用\ref{fig:feature_extraction}）
	\end{figure*}
	
	The ablation experiments consistently demonstrate that each component of the STL-FFT-STFT-TCN-LSTM model significantly contributes to the WVHT prediction performance at Station 41008. Removing any single component leads to increased error, reduced fit quality, or deteriorated stability. This indispensable characteristic validates the superiority of the multi-scale feature fusion strategy. Through the synergistic integration of structural trends, global frequency domain, and dynamic time-frequency domain, the model precisely adapts to complex wave energy data, delivering a robust and efficient marine energy solution for WVHT forecasting.

	\section{Conclusion and outlook}
	
	In recent years, wave energy is of great value due to its characteristics of sustainability, cleanliness, high energy density, and wide distribution. The accurate analysis and prediction of WVHT are key prerequisites for its efficient utilization. Aiming at the complex characteristics of wave energy data such as strong nonlinearity and multi-scale periodic superposition, this study proposes the STL-FFT-STFT-TCN-LSTM model to optimize WVHT prediction. First, the sequence is decomposed into trend items, seasonal items, and residual items through STL. Then, multi-scale frequency features are mined with the help of FFT and STFT. Finally, deep time-series modeling is realized relying on TCN and LSTM. Based on the experimental verification of hourly data from 2019 to 2022 at the near-sea station 41008 and the open-sea station 41047 of NOAA: (1) Compared with other models, whether in the near-sea or open-sea areas, the STL-FFT-STFT-TCN-LSTM model has better accuracy in scenarios of extreme wave height capture and high-frequency noise suppression, showing strong adaptability across sea areas; (2) Ablation experiments further confirm that the multi-scale feature fusion system formed by the collaboration of STL decomposition, FFT, STFT, TCN, and LSTM has indispensable components, which strongly support the model in dealing with complex wave energy characteristics. In conclusion, the STL-FFT-STFT-TCN-LSTM model accurately adapts to the laws of wave energy data through a multi-scale feature fusion strategy. It demonstrates superiority in prediction accuracy and scenario adaptability, providing a robust solution for effective wave height prediction. It is expected to lay a solid technical foundation for wave energy development and utilization, as well as marine environment monitoring services.

	Further experimental and analytical investigations revealed that the pronounced non-linearity and transient abruptness of wave signals, such as the isolated extreme wave heights at offshore Station 41047 influenced by mesoscale eddies, require greater dynamic adaptability in time-frequency resolution. The STFT, constrained by its fixed window length, exhibits limitations in high-frequency temporal localisation and low-frequency resolution, whilst also introducing computational redundancy. Conversely, the discrete wavelet transform (DWT) demonstrates significant advantages through its multiscale decomposition capability. It can precisely capture instantaneous abrupt changes in extreme wave heights, extract long-period patterns, and reduce computational complexity. This aligns highly with the characteristics of wave signals, offering potential for enhanced model accuracy. Future research will retain the strengths of STL and FFT while replacing STFT with DWT to construct an STL-FFT-DWT-TCN-LSTM model, focusing on optimising wavelet basis selection and feature fusion mechanisms. Concurrently, the TCN-LSTM architecture will be enhanced by integrating Seq2Seq structures, employing tailored strategies for different prediction horizons. For instance, short-term forecasts such as 1-hour or 2-hour intervals will utilise a ‘fine-grained input + real-time feature updating’ approach to strengthen high-frequency noise suppression. Conversely, long-term predictions spanning 6 hours or more will incorporate historical error feedback mechanisms to refine input sequences and mitigate cumulative inaccuracies. Furthermore, the data scope is expanded by integrating wave energy device data to establish a coupled model, thereby enhancing the model's practical value in wave energy development projects.

	%% Loading bibliography style file
	%\bibliographystyle{model1-num-names}
	% \bibliographystyle{unsrt}
	% % \bibliographystyle{sn-bibliography}
	
	% % Loading bibliography database
	% \bibliography{cas-refs}

\end{document}